\documentstyle{aipproc}
\input boxedeps
\SetepsfEPSFSpecial 
\HideDisplacementBoxes

\newcommand{\etal}{{\em et al.}}
\newcommand{\ie}{{\em i.e.}}

\newcommand{\kmwe}{\hbox{ kmwe}}
\newcommand{\gevcc}{\hbox{ GeV}\!/\!c^2}
\newcommand{\gev}{\hbox{ GeV}}

\newcommand{\ev}{\hbox{ eV}}
\newcommand{\evcc}{\hbox{ eV}\!/\!c^2}
\newcommand{\mev}{\hbox{ MeV}}
\newcommand{\mevcc}{\hbox{ MeV}\!/\!c^2}
\newcommand{\tev}{\hbox{ TeV}}

\newcommand{\tevcc}{\hbox{ TeV}\!/\!c^2}

\newcommand{\ys}{\hbox{ ys}}
\newcommand{\ps}{\hbox{ ps}}
\newcommand{\s}{\hbox{ s}}

\newcommand{\cm}{\hbox{ cm}}
\newcommand{\mm}{\hbox{ mm}}

\newcommand{\fb}{\hbox{ fb}}
\newcommand{\km}{\hbox{ km}}
\newcommand{\m}{\hbox{ m}}

\newcommand{\eqn}[1]{(\ref{#1})}
\def\half{{\scriptstyle \frac{1}{2}}}     

\def\slashii#1{\setbox0=\hbox{$#1$}             
   \dimen0=\wd0                                 
   \setbox1=\hbox{\sl/} \dimen1=\wd1            
   \ifdim\dimen0>\dimen1                        
      \rlap{\hbox to \dimen0{\hfil\sl/\hfil}}   
      #1                                        
   \else                                        
      \rlap{\hbox to \dimen1{\hfil$#1$\hfil}}   
      \hbox{\sl/}                               
   \fi}                                         %
\def\ltap{\mathop{\raisebox{-.4ex}{\rlap{$\sim$}} 
\raisebox{.4ex}{$<$}}}

\newcommand{\cfrac}[2]{\textstyle \frac{#1}{#2}}

\def\bentarrow{\:\raisebox{1.3ex}{\rlap{$\vert$}}\!\rightarrow}

\def\dk#1#2#3{
	\begin{equation}
	\begin{array}{r c l}
	#1 & \rightarrow & #2 \\
	 & & \bentarrow #3
	\end{array}
	\end{equation}
		}

\def\bothdk#1#2#3#4#5#6{
	\begin{equation}
	\begin{array}{r c l}
	#1 & \rightarrow & #2#3 \\
	 & & \:\raisebox{1.3ex}{\rlap{$\vert$}}\raisebox{-0.5ex}{$\vert$}%
	\phantom{#2}\!\bentarrow #4 \\
	 & & \bentarrow #5
	\end{array}\label{eq:#6}
	\end{equation}
		}

\newcommand{\hepex}[1]{hep-ex/#1}
\newcommand{\hepph}[1]{hep-ph/#1}
\newcommand{\hepth}[1]{hep-th/#1}

\newcommand{\astro}[1]{astro-ph/#1}
\def\prl#1#2#3{\frenchspacing{\it Phys. Rev. Lett. }{\bf #1}, #2 (19#3)}
\def\prll#1#2#3{\frenchspacing{\it Phys. Rev. Lett. }{\bf #1}, #2 (#3)}
\def\pr#1#2#3{\frenchspacing{\it Phys. Rev. D}{\bf #1}, #2 (19#3)}

\def\prc#1#2#3{\frenchspacing{\it Phys. Rev. C}{\bf #1}, #2 (19#3)}
\def\prev#1#2#3{\frenchspacing{\it Phys. Rev. }{\bf #1}, #2 (19#3)}
\def\pl#1#2#3{\frenchspacing{\it Phys. Lett. }{\bf #1}, #2 (19#3)}
\def\np#1#2#3{\frenchspacing{\it Nucl. Phys. }{\bf #1}, #2 (19#3)}
\def\rmp#1#2#3{\frenchspacing{\it Rev. Mod. Phys. }{\bf #1}, #2 (19#3)}

\def\ib#1#2#3{{\bf #1}, #2 (19#3)}

\def\phystoday#1#2#3#4{\frenchspacing{\it Phys. Today\/ }{\bf #1}, #2 
(\ifcase#3\or January\or 
         February\or March\or April\or May\or June\or July\or August\or 
         September\or October\or November\or December\fi, 19#4)}

\begin{document}

\title{Perspectives in High-Energy Physics\footnote{Review lecture 
given at the ICFA Instrumentation School, Istanbul, Turkey, June 30, 
1999.}}
\author{Chris Quigg\thanks{Internet address: 
\textsf{quigg@fnal.gov.}}} \address{Theoretical Physics Department \\ 
Fermi National Accelerator Laboratory~\thanks{Fermilab is operated by 
Universities Research Association Inc.\ under Contract No.\ 
DE-AC02-76CH03000 with the United States Department of Energy.\hfill 
\fbox{\textsf{FERMILAB--Conf--00/041--T}}}\\ P.O.  Box 500, Batavia, Illinois 
60510 USA}

\lefthead{\thepage \hfill \textsf{FERMILAB--Conf--00/041--T}}
\righthead{\textsf{FERMILAB--Conf--00/041--T} \hfill \thepage}
\maketitle

\begin{abstract}
I sketch some pressing questions in several active areas of particle physics 
and outline the challenges they present for the design and operation 
of detectors.
\end{abstract}

\section*{Introduction}
My assignment at the 1999 ICFA Instrumentation School is to survey 
some current developments in particle physics, and to describe the 
kinds of experiments we would like to do in the near future and 
illustrate the demands our desires place on detectors and data 
analysis.  Like any active science, particle physics is in a state of 
continual renewal.  Many of the subjects that seem most 
fascinating and most promising today simply did not exist as recently 
as twenty-five years ago.  Other topics that have preoccupied
physicists for many years have been reshaped by recent discoveries and
insights, and transformed by new techniques in accelerator science and
detector technology.  To provide some context for the courses and
laboratories at this school, I have chosen three topics that are of
high scientific interest, and that place very different demands on
instrumental techniques.  I hope that you will begin to see the
breadth of opportunities in particle physics, and that you will also
look beyond the domain of particle physics for opportunities to apply
the lessons you learn here in Istanbul.

I begin with the remarkable neutrino, a subatomic particle that 
our instruments must be able to detect both by its presence and by its 
absence, depending on the circumstances.  In particular, I note the 
interest in observing and characterizing neutrino oscillations in 
order to determine the properties of the neutrinos, and describe some 
of the new instruments contemplated to do that.  Then I will talk 
about physics at the high-energy frontier, focusing on what we hope to 
learn from the top quark in the next set of experiments at Fermilab's 
Tevatron Collider.  Third, I will tell you some of the ways we hope to 
explore the landscape of spacetime, and explore some of 
the signs we might find that the three-plus-one dimensions of ordinary 
experience are not the whole story.  My treatment of all of these will 
be schematic; in a single lecture, my intent is to raise questions and 
present a wide range of challenges and opportunities.

\section*{Neutrino Puzzles}
Neutrinos are tiny subatomic particles that carry no electric charge, 
have (almost) no mass, move (nearly) at the speed of light, and hardly 
interact at all.  They are among the most abundant particles in 
the Universe.  As you listen to this lecture, inside your body are 
more than 10 million ($10^{7}$) relic neutrinos left over from the Big 
Bang.  Each second, some $10^{14}$ neutrinos made in the Sun pass 
through you.  In one tick of the clock, about a thousand neutrinos made by 
cosmic-ray interactions in Earth's atmosphere traverse your body.  
Other neutrinos reach us from natural sources, including radioactive 
decays of elements inside the Earth, and artificial sources, such as 
nuclear reactors.

Our awareness of neutrinos started with a puzzle in 1914 that led to 
an idea in 1930 that was confirmed by an experiment in 1956.  Today, 
neutrinos have become an important tool for particle physics and 
astrophysics, and fascinating objects of study that may yield 
important new clues about the basic laws of Nature.

\subsection*{The First Neutrino Puzzle} 
\begin{figure}[tb] 
\centerline{\BoxedEPSF{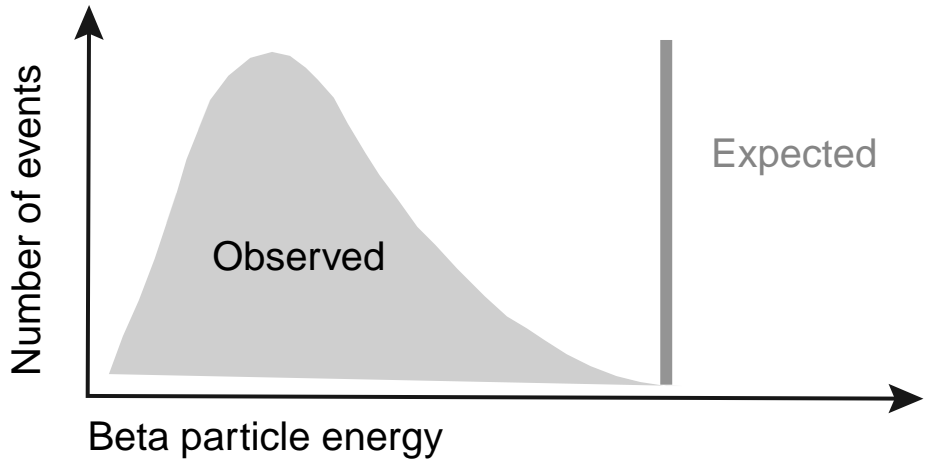  scaled 1500}}
\vspace{10pt}
\caption{Expectations and reality for the beta decay spectrum.}
\label{fig:betaspec}
\end{figure}

There was a neutrino puzzle even before physicists knew there was a 
neutrino.  Natural and artificial radioactivity includes nuclear beta 
($\beta$) decay, observed as
\begin{equation}
	^{A}{\mathrm{Z}} \to\ ^{A}({\mathrm{Z+1}}) + \beta^{-}\; ,
	\label{eq:betadk}
\end{equation}
where $\beta^{-}$ is the old-fashioned name for an electron and
$^{A}{\mathrm{Z}}$ stands for the nucleus with $Z$ protons and $A-Z$
neutrons.  Examples are tritium $\beta$ decay,
\begin{equation}
	^{3}\mathrm{H}_{1} \to\ ^{3}\mathrm{He}_{2} + \beta^{-}\; ,
\end{equation}
neutron $\beta$ decay,
\begin{equation}
	n \to p + \beta^{-}\; ,
\end{equation}
and $\beta$ decay of Lead-214,
\begin{equation}
	^{214}\mathrm{Pb}_{82} \to\ ^{214}\mathrm{Bi}_{83} + \beta^{-}\; .
\end{equation}
For two-body decays, the Principle of Conservation of Energy \& 
Momentum says that the $\beta$ particle, or electron, should have a 
definite energy, indicated by the spike in Figure \ref{fig:betaspec}.  
What was observed was very different: in 1914, James Chadwick (later 
to discover the neutron) showed conclusively \cite{chadwick} that in 
the decay of Radium B and C ($^{214}$Pb and $^{214}$Bi), the $\beta$ 
energy follows a continuous spectrum, as shown in Figure 
\ref{fig:betaspec}.

\begin{figure}[b!] 
\centerline{\BoxedEPSF{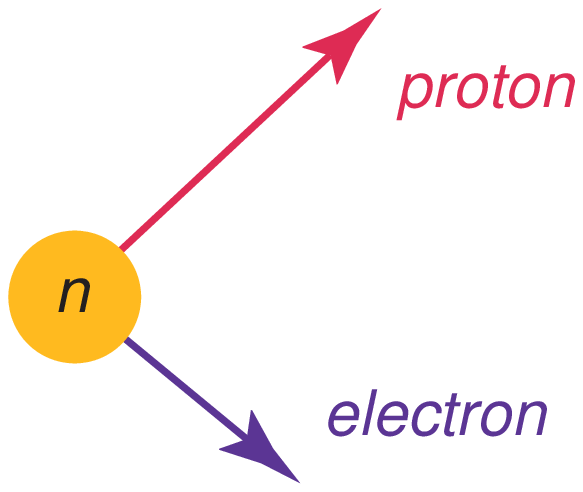  scaled 600}\qquad
\BoxedEPSF{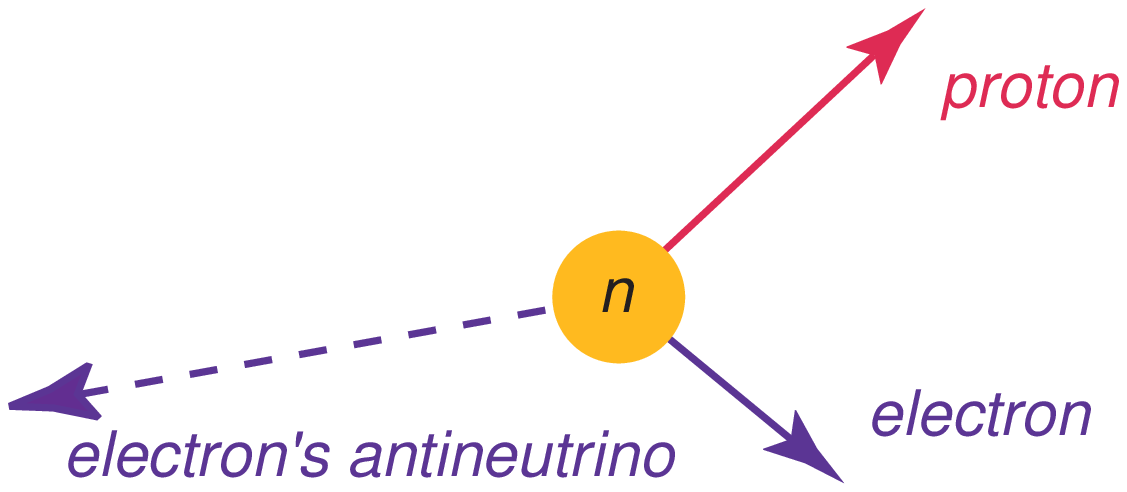  scaled 600}}
\vspace{10pt}
\caption{\textit{(Left) }Apparent nonconservation of energy and 
momentum in neutron beta decay; \textit{(Right) }Pauli's solution to 
the nonconservation of energy and momentum: a nearly massless, 
neutral, penetrating particle that we know as $\bar{\nu}_{e}$.}
\label{fig:Enocons}
\end{figure}

How could we account for this completely unexpected behavior?  Might 
it mean that energy and momentum are not uniformly conserved in 
subatomic events?\footnote{Niels Bohr was willing to consider this 
possibility.} Although Chadwick did not discover the neutron until 
1932, we can use a little chronological license to sharpen our puzzle 
by considering a cartoon of neutron $\beta$ decay.  The continuous 
$\beta$ spectrum means that, in general, the products of the decay of 
a stationary neutron will not have balanced momenta (zero net 
momentum), as shown in the left-hand frame in Figure 
\ref{fig:Enocons}.  For the products of a system at rest to drift off 
in some direction flies in the face of physical intuition, though we 
have to concede that at the time, physicists' intuition was still 
largely derived from macroscopic experience.

\subsection*{The Neutrino Conjectured and Observed} 
The $\beta$-decay energy crisis tormented physicists for years.  On 
December 4, 1930, Wolfgang Pauli addressed an open letter\footnote{Pauli's letter (in the 
original German) is reproduced in Ref.\ \cite{pauliworks}. For an 
English translation, see pp.~127-8 of Ref.\ \cite{bram}.  It begins, 
``Dear Radioactive Ladies and Gentlemen, I have hit upon a desperate 
remedy regarding \ldots the continuous $\beta$-spectrum \ldots'' Pauli 
concluded, ``For the moment I dare not publish anything about this 
idea and address myself confidentially first to you.  \ldots I admit 
that my way out may seem rather improbable \textit{a priori} \ldots .  
Nevertheless, if you don't play you can't win \ldots .  Therefore, 
Dear Radioactives, test and judge.'' Pauli's neutrino, together with 
the discovery of the neutron, also resolved a vexing nuclear 
spin-and-statistics problem.} to a meeting on radioactivity in 
T\"{u}bingen.  Pauli could not attend in person because his presence 
at a student ball in Zurich was ``indispensable.'' In his letter, 
Pauli advanced the outlandish idea of a new, very penetrating, neutral 
particle of vanishingly small mass.  Because Pauli's new particle 
interacted very feebly with matter, it would escape undetected from 
any known apparatus, taking with it some energy, which would seemingly 
be lost.  The balance of energy and momentum would be restored, as 
shown in the right-hand frame of Figure \ref{fig:Enocons}, by the 
particle we now know as the electron's antineutrino.  The proper decay 
scheme for the neutron is thus
\begin{equation}
	n \to\ p + \beta^{-} + \bar{\nu} \; .
	\label{eq:betadknu}
\end{equation}

Pauli's new particle was indeed a ``desperate remedy,'' but it was, 
in its way, very conservative, for it preserved the principle of 
energy and momentum conservation and with it the notion that the laws 
of physics are invariant under translations in space and time.  The 
hypothesis fit the facts; after the discovery of the neutron in 1932, 
Fermi named the new particle the neutrino, to distinguish it from the 
neutron, and constructed his four-fermion theory of the weak interaction.  
Experimental confirmation of Pauli's neutrino had to wait for dramatic 
advances in technology.

Detecting a particle as penetrating as the neutrino required a large 
target and a copious source of neutrinos.  In 1953, Clyde Cowan and 
Fred Reines \cite{cowan} used the intense beam of antineutrinos from a 
fission reactor
\begin{equation}
	^{A}{\mathrm{Z}} \to\ ^{A}({\mathrm{Z+1}}) + 
	\beta^{-} + \bar{\nu}\; ,
	\label{eq:betapdk}
\end{equation}
and a heavy target ($10.7~\mathrm{ft}^{3}$ of liquid scintillator) containing 
about $10^{28}$ protons to detect the reaction
\begin{equation}
	\bar{\nu} + p \to\ e^{+} + n \; .
	\label{eq:invbdk}
\end{equation}
Initial runs at the Hanford Engineering Works were suggestive but inconclusive.  
Moving their apparatus to the stronger fission neutrino source at the 
Savannah River nuclear plant, Cowan and Reines and their team made the 
definitive observation of inverse $\beta$ decay in 1956 \cite{reines}.

\subsection*{Three Families of Leptons}
In addition to the electron ($e$) and its neutrino ($\nu_{e}$), we now 
recognize two other pairs of pointlike, spin-$\cfrac{1}{2}$ particles 
that are not affected by the strong interaction:
	\begin{equation}
		\left(
		\begin{array}{c}
			\nu_{e}  \\
			e^{-}
		\end{array}
		 \right)_{L} \;\;\;\;\;\;
		\left(
		\begin{array}{c}
			\nu_{\mu}  \\
			\mu^{-}
		\end{array}
		 \right)_{L} \;\;\;\;\;\;
		\left(
		\begin{array}{c}
			\nu_{\tau}  \\
			\tau^{-}
		\end{array}
		 \right)_{L}\; .		 
	\end{equation}
These particles are known as leptons, from the Greek 
$\lambda\epsilon\pi\tau \acute{o}\varsigma$ = thin, inspired by the 
small mass of the electron, muon, and their neutrinos compared with 
the mass of the proton and neutron, the lightest of the baryons (from 
the Greek $\beta\alpha\rho\acute{\upsilon}\varsigma$ = 
heavy).\footnote{We would not choose the same names today, knowing 
that the tau lepton is twice as massive as the proton \ldots}

The muon neutrino is created in charged pion decay,
\begin{equation}
    \pi^{+} \to \mu^{+}\nu_{\mu} \quad\hbox{and}\quad
    \pi^{-} \to \mu^{-}\bar{\nu}_{\mu}\; ,
    \label{eq:pitomu}
\end{equation}
and neutrino beams produced at accelerators are overwhelmingly muon 
neutrinos.  The two-neutrino experiment carried out at Brookhaven 
National Lab in the early 1960s \cite{2nu} demonstrated that the 
neutrinos produced in pion decays do not initiate inverse $\beta$ 
decay, so that $\nu_{\mu}$ is distinct from $\nu_{e}$.

The \textsc{donut} (\textsc{d}irect \textsc{o}bservation of
\textsc{nu}-\textsc{t}au) Experiment \cite{donut} under analysis at
Fermilab is a three-neutrino experiment.  Using a prompt neutrino beam 
in which decays of the charmed-strange meson
\dk{D_{s}^{+}}{\tau^{+}\nu_{\tau}}{\bar{\nu}_{\tau}+\hbox{anything}}
provide the $\nu_{\tau}$ source, the experimenters aim to observe the 
reaction
\begin{equation}
    \nu_{\tau}N \to \tau + \hbox{anything.}
    \label{eq:tauapp}
\end{equation}
I expect to see results from \textsc{donut} in the year 2000.

\subsection*{Neutrino Beams and Detectors}
\begin{figure}[tb] 
\centerline{\BoxedEPSF{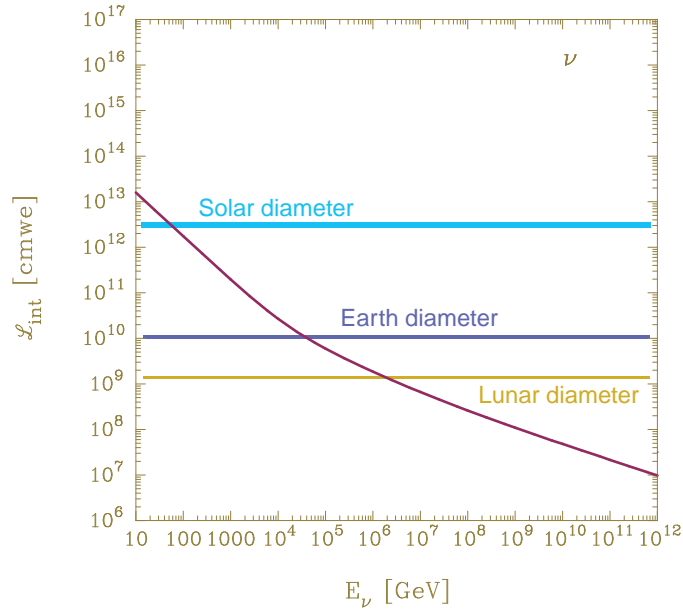  scaled 1000}}
\vspace{10pt}
\caption{Interaction lengths for neutrino interactions on an isoscalar
nucleon target (from Ref.\ {\protect\cite{gqrs}}).}
\label{fig:Lint}
\end{figure}

Neutrinos traverse vast amounts of material.  The fission 
antineutrinos detected by Cowan, Reines, and their collaborators have 
an inverse $\beta$-decay cross section $\sigma(\bar{\nu}_{e}p 
\rightarrow e^{+}n) \approx 10^{-43}\cm^{2}$.  Accordingly, their 
interaction length, 
\begin{equation}
    {\mathcal{L}}_{\textrm{int}} = \frac{1}{\sigma(\bar{\nu}_{e}p \rightarrow e^{+}n) 
    N_{A} (Z/A) \bar{\rho}}\; , 
    \label{eq:lintdef}
\end{equation}
where $N_{A} = 6.022 \times 10^{23}\cm^{-3}$ is Avogadro's number, 
$\bar{\rho}$ is the specific gravity of the target, and $Z/A$ is the 
proton fraction of the target nucleons, is very long.  It corresponds 
to about $1.7 \times 10^{19}\cm$ of water, or a column density of 
$1.7 \times 10^{10}$ kilotonnes per square centimeter.  On average, a 
fission neutrino would traverse more than four light-years of 
lead---the distance from Earth to $\alpha$-Centauri---before 
interacting.  At higher energies, the cross section for the 
inclusive reaction $\bar{\nu}_{e} p \to e^{+} + \hbox{anything}$ 
grows---at first $\propto E_{\nu}$, then more slowly---so the 
interaction length decreases.  The high-energy dependence of the $\nu 
N$ interaction length is shown in Figure \ref{fig:Lint}.
The interaction length of a 100-GeV neutrino is approximately 25 
million kilometers of water, or about 230 Earth diameters.  If you 
should stand in Fermilab's neutrino beam, only one neutrino in 
$10^{11}$ will interact in your body.

What are the consequences of the great interaction length?
First, there is the missing-energy signature for neutrinos.  Second, 
there is the difficulty of detecting neutrino interactions.  
Third---because the other known particles all interact more than 
neutrinos do---we have the possibility of preparing filtered neutrino 
beams, which we shall discuss in a moment.  Fourth, because neutrinos 
can penetrate great columns of matter, whereas electromagnetic 
radiation is blocked by a few hundred grams of material, it is 
appealing to consider the promise of neutrino astronomy for peering 
into the hearts of dense structures or looking back at the state of 
the universe before recombination of ions and electrons into neutral 
atoms.

At high-energy accelerators, neutrinos are tertiary products of 
collisions
\begin{equation}
	p + \hbox{Target} \rightarrow \hbox{many }\pi,\;K\;\;,
\end{equation}
followed by the decays
\begin{equation}
	\pi \to \mu \nu_{\mu}, \;\;\; K \to \mu \nu_{\mu}\; ,
\end{equation}
in an evacuated decay space up to a kilometer long.  The decay region 
is followed by an earthen shield, perhaps made denser by the 
inclusion of blocks of iron or steel, that absorbs photons, surviving 
mesons, and other hadrons, and ranges out many of the muons.  At 
Fermilab, the total length of the neutrino beam line is about $3\km$.

In the most recent study of deeply inelastic neutrino scattering at
Fermilab, 800-GeV protons from the Tevatron deliver $10^{10}$
neutrinos in 5 pings over 2.5 seconds each minute, spread over the
$100\hbox{ ft}^{2}$ face of the NuTeV Detector.  The detector is made
up of 690~T of iron, scintillator, and drift chambers.  Events are
studied from a ``fiducial volume'' of 390~T. The $10^{10}$ neutrinos
per minute produce about 10 -- 20 events that are recorded for
off-line analysis.  Experiments of this kind give us our best look at
the interior of the proton, and reveal its quark structure in
exquisite detail \cite{nutev}.

\subsection*{Neutrino Mass}
\textit{Neutrinos are very light.} 
No one has ever weighed a neutrino.  The best kinematical 
determinations set upper bounds \cite{pdg} on the dominant 
neutrino species emitted in nuclear beta decay ($m_{\nu_{e}} \ltap 
15\evcc$), charged-pion decay ($m_{\nu_{\mu}} < 0.19\mevcc$ at 
90\% CL), and $\tau$-lepton decay ($m_{\nu_{\tau}} < 18.2\mevcc$ at
95\% CL).  Although there are prospects\footnote{Some of these
prospects were reviewed at $\nu$Fact '99 in Lyon by Alvaro De
R\'ujula, Ref.  \cite{alvaro}.} for improving these bounds---and the
measurement of a nonzero mass would constitute a real discovery---they
are sufficiently large that it is of interest to consider indirect
(nonkinematic) constraints from other quarters.

If neutrino lifetimes are greater than the age of the Universe, the 
requirement that neutrino relics from the Big Bang not overclose the 
Universe leads to a constraint on the sum of neutrino masses.  For 
relatively light neutrinos ($m_{\nu} \ltap\hbox{a few}\mevcc$), the 
total mass in neutrinos,
\begin{equation}
	m_{\mathrm{tot}} = \sum_{i}\half g_{i} m_{\nu_{i}} \; ,
	\label{totmass}
\end{equation}
where $g_{i}$ is the number of spin degrees of freedom of $\nu_{i}$ 
plus $\bar{\nu}_{i}$, sets the scale of the neutrino contribution to 
the mass density of the Universe, $\varrho_{\nu} = m_{\mathrm{tot}}n_{\nu} 
\approx 112\,m_{\mathrm{tot}}\cm^{-3}$.  If we measure 
$\varrho_{\nu}$ as a fraction of the critical density to close the 
Universe, $\varrho_{c} = 1.05 \times 10^{4}\,h^{2}\evcc \cm^{-3}$, 
where $h$ is the reduced Hubble parameter, then
\begin{equation}
	\Omega_{\nu} \equiv \frac{\varrho_{\nu}}{\varrho_{c}} = 
	\frac{m_{\mathrm{tot}}}{94h^{2}\evcc} \; .
	\label{omgnu}
\end{equation}
An assumed bound on $\Omega_{\nu}h^{2}$ then implies a bound on 
$m_{\mathrm{tot}}$.  A very conservative bound results from the 
assumption that $\Omega_{\nu}h^{2} < 1$: it is that $m_{\mathrm{tot}} < 
94\evcc$.

Recent observations\footnote{For a review and interpretation of recent 
observations, see Ref.\ \cite{costri}.} suggest that the total matter 
density is considerably smaller than the critical density, so that 
$\Omega_{m} \approx 0.3$.  If we fix $\Omega_{\nu} < \Omega_{m}$ and 
choose the plausible value $h^{2} = 0.5 \pm 0.15$, then we arrive at 
the still generous upper bound $m_{\mathrm{tot}} \ltap 19\evcc$.  
Taking into account the best (and model-dependent) information about 
the hot- and cold-dark-matter cocktail \cite{mst} needed to reproduce 
the observed fluctuations in the cosmic microwave background, it seems 
likely that cosmology limits $m_{\mathrm{tot}} \ltap \hbox{a 
few}\evcc$.  It is worth remarking that the cosmological desire for 
hot dark matter has been on the wane.

If neutrinos were exactly massless, neutrino physics would be simple.  
There would be no pattern of masses to explain, no neutrino decays, no 
mixing among lepton generations, and no neutrino oscillations.  The 
only question would be Why?

If neutrinos do have masses, the electroweak theory has more 
unexplained parameters: the neutrino masses, three mixing angles, and 
a \textsf{CP}-violating phase.  Neutrinos may decay, and neutrinos 
can oscillate from one electroweak species to another.

\subsection*{Neutrino Oscillations\footnote{For a review of the 
essentials, see {\protect\cite{wick}}.}}
In the quantum world, particles are waves.  If neutrinos $\nu_{1}, 
\nu_{2}, \ldots$ have different masses $m_{1}, m_{2}, \ldots$ ,
each neutrino flavor may be a mixture of different masses.  Let us 
consider two species for simplicity, and take 
\begin{equation}
\left( 
	\begin{array}{c}
	\nu_{e}  \\
	\nu_{\mu}
	\end{array}
\right) = \left( 
	\begin{array}{cc}
	\cos\theta & \sin\theta  \\
	-\sin\theta & \cos\theta
	\end{array}
\right) \left( 
	\begin{array}{c}
	\nu_{1}\\
	\nu_{2}
	\end{array}
\right)\; .
\end{equation}
If neutrinos are emitted with a definite momentum $p$, the wave 
functions corresponding to the two mass eigenstates evolve with 
different frequencies.  As a consequence, a beam born as pure 
$\nu_{\mu}$ may evolve a $\nu_{e}$ component with time.  If the 
neutrino momentum is large compared with the neutrino masses, $p \gg 
m_{i}$, then the probability for a $\nu_{e}$ component to develop in a 
$\nu_{\mu}$ beam after a time $t$ is
\begin{equation}
P_{\nu_{e}\leftarrow\nu_{\mu}}(t) = \sin^{2}2\theta \sin^{2}\left(
\frac{\Delta m^{2}\,t}{4p}\right)\; .
\end{equation}
Measuring the propagation distance $L = ct$, approximating the 
neutrino energy as $E \approx pc$, and using the conversion factor 
$\hbar c \approx 1.97 \times 10^{-13}\mev\m$, we can re-express
\begin{equation}
    \sin^{2}\left(\frac{\Delta m^{2}t}{4p}\right) \approx
    \sin^{2}\left(1.27 \frac{\Delta m^{2}}{1\ev^{2}} \cdot \frac{L}{1\km} 
    \cdot \frac{1\gev}{E}\right)\; .
    \label{eq:metamorph}
\end{equation}
Arising as it does from a slight frequency mismatch, neutrino flavor
oscillation is analogous to the beat-frequency phenomenon observed 
when two tuning forks with \textit{almost} the same pitch are sounded 
at the same time.  If the forks are sounded individually, producing 
waves \newline
\phantom{rufus}\newline
\centerline{\BoxedEPSF{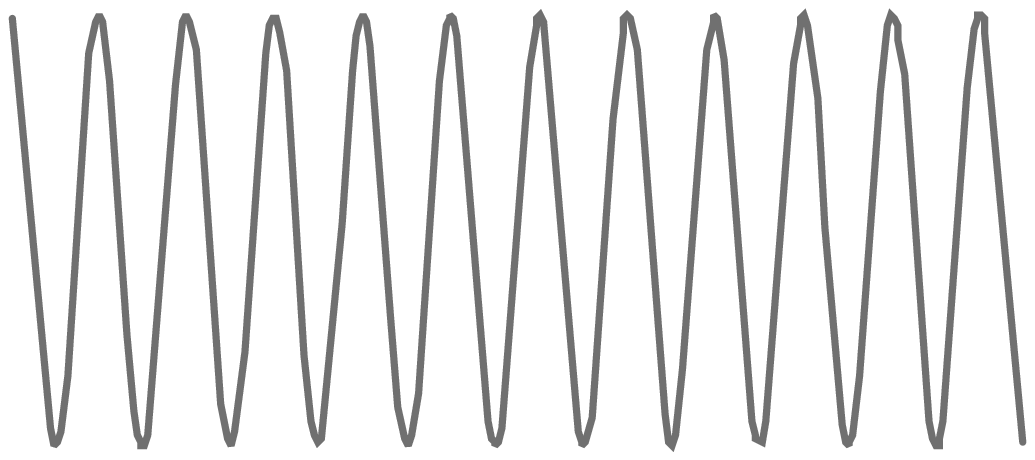  scaled 250} or
\BoxedEPSF{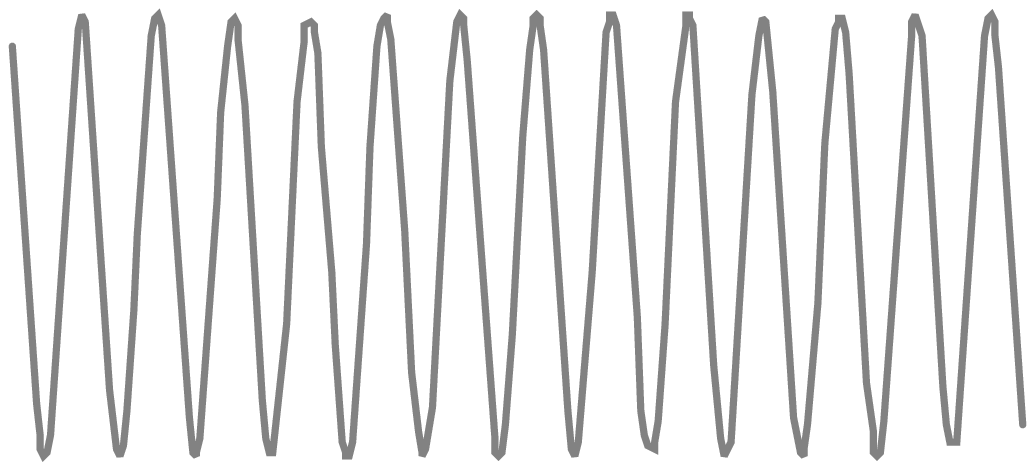  scaled 250} ,}
\phantom{rufus}\newline
we might find it difficult to distinguish the pitches.  But when we 
sound them together, the sound intensity \newline
\centerline{$\left|\BoxedEPSF{sin6x2.eps  scaled 250} + 
\BoxedEPSF{sin7x2.eps  scaled 250}\right|^{2} = 
\;\BoxedEPSF{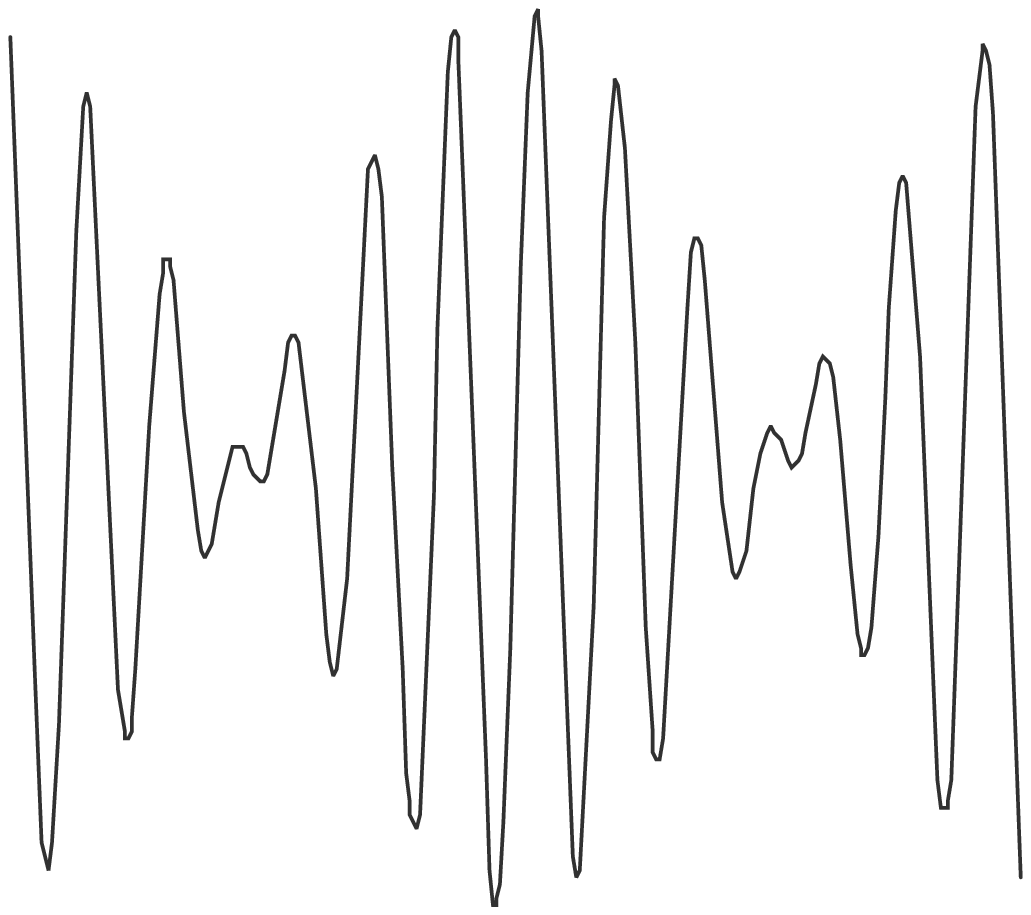  scaled 250}$}
\phantom{rufus}\newline
swells and fades periodically because the two sound waves 
are different, reflecting the physical difference between the two 
tuning forks.

The probability that a neutrino born as $\nu_{\mu}$ remain a 
$\nu_{\mu}$ at distance $L$ is
\begin{equation}
	P_{\nu_{\mu}\leftarrow\nu_{\mu}}(L) = 
1 - \sin^{2}2\theta \sin^{2}\left(1.27 \frac{\Delta m^{2}}{1\ev^{2}} 
\cdot \frac{L}{1\km} \cdot \frac{1\gev}{E}\right)\; .
\end{equation}

\begin{figure}[tb] 
\centerline{\BoxedEPSF{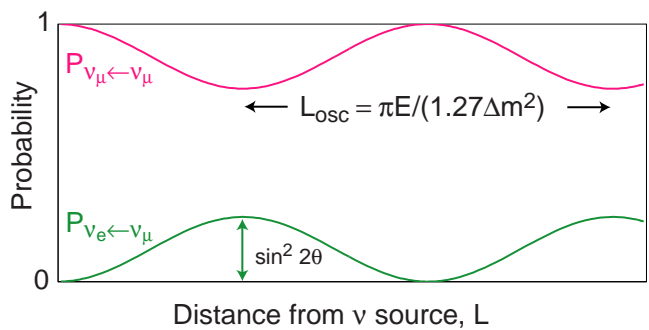  scaled 750}}
\vspace{10pt}
\caption{Evolution of a $\nu_{\mu}$ beam of energy $E$ depends on the 
intrinsic parameters $\Delta m^{2}$ and $\sin^{2}2\theta$, and on the 
experimental conditions characterized by $L$ and $E$.}
\label{fig:oscparam}
\end{figure}

The probability for a $\nu_{\mu}$ to metamorphose into a $\nu_{e}$,
\begin{equation}
P_{\nu_{e}\leftarrow\nu_{\mu}} = \sin^{2}2\theta \sin^{2}\left(1.27 
\frac{\Delta m^{2}}{1\ev^{2}} \cdot \frac{L}{1\km} \cdot 
\frac{1\gev}{E}\right)\; ,
\end{equation}
depends on two parameters related to experimental conditions: $L$, the 
distance from the neutrino source to the detector, and $E$, the 
neutrino energy.  It also depends on two fundamental neutrino 
parameters: the difference of masses squared, $\Delta m^{2} = 
m_{1}^{2} - m_{2}^{2}$, and the neutrino mixing parameter, 
$\sin^{2}2\theta$.  The amplitude of the probability oscillations is 
given by $\sin^{2}2\theta$, as shown in Figure
\ref{fig:oscparam}; the wavelength of the oscillations is 
\begin{equation}
    L_{\mathrm{osc}} = 
    \frac{\pi}{1.27}\cdot\frac{E}{1\gev}\cdot\frac{1\ev^{2}}{\Delta 
    m^{2}}\km = 2.48\, \frac{E}{1\gev}\cdot\frac{1\ev^{2}}{\Delta 
    m^{2}}\km\; ,
    \label{eq:wavelength}
\end{equation}
and the oscillation probability is greatest at a distance 
$L_{\mathrm{max}}=(k+\cfrac{1}{2})L_{\mathrm{osc}}$, where $k$ is an 
integer.

Many experiments have now used natural sources of neutrinos, neutrino 
radiation from fission reactors, and neutrino beams generated in 
particle accelerators to look for evidence of neutrino
oscillation.\footnote{For summaries of the current evidence about
neutrino oscillations, see Ref.\ \cite{janetc}.}

\begin{figure}[tb] 
\centerline{\BoxedEPSF{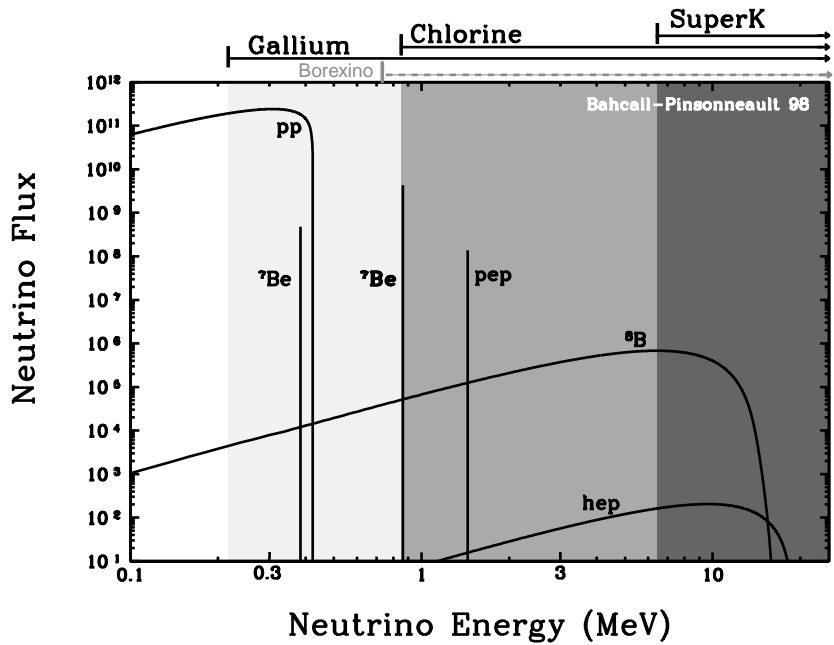  scaled 600}}
\vspace{10pt}
\caption{The spectrum of neutrinos from the $pp$ chain predicted by 
the standard solar model.  The neutrino fluxes from continuum sources 
($pp$ and $^{8}$B) are given in units of number per cm$^{2}$ per 
second per MeV at one astronomical unit.  The line fluxes ($pep$ and 
$^{7}$Be) are given in number per cm$^{2}$ per second.  The detection 
thresholds for several solar neutrino experiments are indicated at the 
top of the figure.  The figure is from Ref.\ {\protect\cite{jnbapj}}, 
updated using the data from Ref.\ {\protect\cite{jnb98}}.}
\label{fig:sunjb}
\end{figure}

The nuclear burning that powers the Sun produces neutrinos as well as 
light and heat \cite{jnbna}.  Overall, a network of reactions we may 
summarize as 
\begin{equation}
	4p \rightarrow\ ^{4}\mathrm{He} + 2e^{+} + 2\nu_{e} + 25\mev
	\label{eq:nucburn}
\end{equation}
leads to the spectrum of neutrinos\footnote{The standard solar model 
is described in Ref.\ \cite{ssm}.  A wealth of information 
is available on John Bahcall's home page at 
\textsf{http://www.sns.ias.edu/\~{}jnb}.} shown in Figure \ref{fig:sunjb}. 
Because solar neutrinos interact so feebly, they can only be detected 
in a very massive target and detector.  The Super-Kamiokande 
Detector in Japan consists of $50\,000$ tonnes of pure water viewed by 
$11\,000$ photomultiplier tubes to detect Cherenkov light.  It is 
sited $1\km$ under a mountain, under $3\kmwe$.  The great advantage of 
the Super-K detector is that it detects neutrino interactions 
$\nu_{e}+n \to p + e^{-}$ in real time, and determines the neutrino 
direction from the electron direction.  Super-K has demonstrated that 
the brightest object in the neutrino sky is the Sun, which proves that 
nuclear fusion powers our star.  The disadvantage of the 
water-Cherenkov technique (see Figure \ref{fig:sunjb}) is that it is 
only sensitive to the highest-energy solar neutrinos.

Five solar-neutrino experiments report deficits with respect to the
predictions of the standard solar model: Kamiokande and
Super-Kamiokande using water-Cherenkov techniques, SAGE and GALLEX
using chemical recovery of germanium produced in neutrino interactions
with gallium, and Homestake using radiochemical separation of argon
produced in neutrino interactions with chlorine.  These results
suggest the oscillation $\nu_{e} \rightarrow \nu_{x}$.

\begin{figure}[tb] 
\centerline{\BoxedEPSF{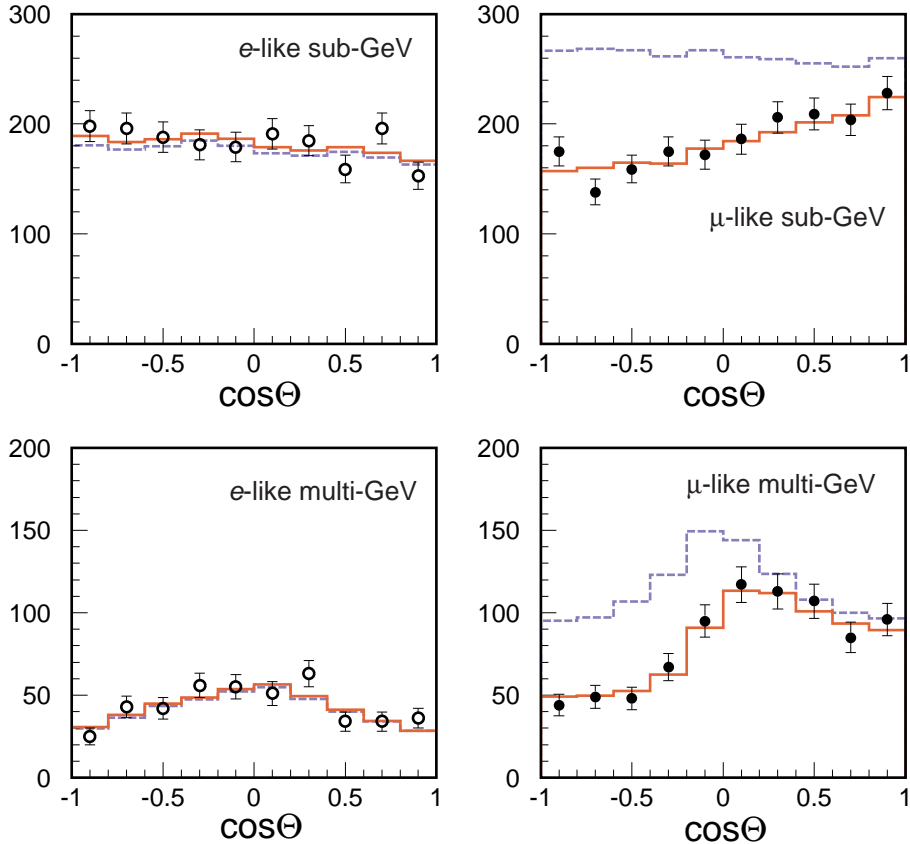  scaled 1000}}
\vspace{10pt}
\caption{Comparison of Monte Carlo simulations (dashed curves) and 
measurements for events initiated in the Super-Kamiokande detector by 
cosmic-ray $\nu_{e}+\bar{\nu}_{e}$ ($\circ$) and 
$\nu_{\mu}+\bar{\nu}_{\mu}$ ($\bullet$).  The solid lines show the 
effects of $\nu_{\mu}\to \nu_{\tau}$ oscillations.
}
\label{fig:atmo}
\end{figure}

Cosmic rays that interact in Earth's atmosphere produce neutrinos in
the decays of pions, kaons, and muons, in the approximate proportions
$\nu_{\mu}:\bar{\nu}_{\mu}:\nu_{e}:\bar{\nu}_{e}::2:2:1:1$.  Five
atmospheric-neutrino experiments report anomalies in the arrival of
muon neutrinos: Kamiokande, IMB, and Super-Kamiokande using
water-Cherenkov techniques, and Soudan II and MACRO using sampling
calorimetry.  The most striking result is the zenith-angle dependence
of the $\nu_{\mu}$ rate reported last year by Super-K
\cite{SKatm,SKLyon}, which is shown in Figure \ref{fig:atmo}.  The 
electron-like events follow the Monte Carlo simulation, but the 
muon-like events exhibit a deficit that is most pronounced for 
upward-coming neutrinos, \ie, those for which the flight path is 
longest, up to $13\,000\km$.  These results suggest the oscillation $\nu_{\mu}
\rightarrow \nu_{\tau}\hbox{ or }\nu_{s}$.  Auxiliary information 
disfavors the sterile-neutrino ($\nu_{s}$) interpretation.

The atmospheric- and solar-neutrino anomalies are both disappearance
phenomena.  A single experiment reports the appearance of neutrinos
that would not be seen in the absence of neutrino oscillations.  The
LSND experiment \cite{LSND} reports the observation of
$\bar{\nu}_{e}$-like events is what should be an essentially pure
$\bar{\nu}_{\mu}$ beam produced at the Los Alamos Meson Physics
Facility, suggesting the oscillation $\bar{\nu}_{\mu} \rightarrow
\bar{\nu}_{e}$.  This result has not yet been reproduced by any other
experiment.

A host of experiments have failed to turn up evidence for neutrino
oscillations in the regimes of their sensitivity.  These results limit
neutrino mass-squared differences and mixing angles.  In more than a
few cases, positive and negative claims are in conflict, or at least
face off against each other.  Over the next five years, many
experiments will seek to verify, further quantify, and extend these
claims.

\subsection*{The Ultimate Neutrino Source?}
Our colleagues working to assess the feasibility of very-high-energy 
muon colliders have given us the courage to think that 
it may be possible, not too many years in the future, to accumulate 
$10^{20\mathrm{ - }21}$ (or even $10^{22}$) muons per year.  It is 
very exciting to think of the possibilities that millimoles of muons 
would raise for studies in fundamental physics.

From the perspective of a muon collider, the 2.2-$\mu$s lifetime of 
the muon presents a formidable challenge.  But if the challenge of 
producing, capturing, storing, and replenishing many unstable muons 
can be met, the decays
\begin{equation}
	\mu^{-}  \rightarrow  e^{-}\nu_{\mu}\bar{\nu}_{e}\; , \qquad 
	\mu^{+}  \rightarrow  e^{+}\bar{\nu}_{\mu}\nu_{e}
	\label{mumpdk}
\end{equation}
offer delicious possibilities for the study of neutrino interactions
and neutrino properties \cite{geer,abp,bgw}.  We would have at
our disposal for the first time not only neutrino beams of
unprecedented intensity, but also controlled beams rich in high-energy
electron neutrinos.  In a \textit{Neutrino Factory}, the composition
and spectra of intense neutrino beams will be determined by the
charge, momentum, and polarization of the stored muons.  The beam from
a $\mu^{+}$ storage ring contains $\bar{\nu}_{\mu}$ and $\nu_{e}$, but
no $\nu_{\mu}$, $\bar{\nu}_{e}$, $\nu_{\tau}$, or $\bar{\nu}_{\tau}$. 
The neutrino spectra are given by
\begin{equation}
	\frac{d^{2}N_{\bar{\nu}_{\mu}}}{dxd\Omega} = 
	\frac{x^{2}}{2\pi}[(3-2x)-(1-2x)\cos\theta]\;,\quad 
	\cos\theta = \hat{p}_{\nu}\cdot 
	\hat{s}_{\mu}\;,
\end{equation}
and
\begin{equation}
	\frac{d^{2}N_{\nu_{e}}}{dxd\Omega} = 
	\frac{3x^{2}}{\pi}[(1-x)-(1-x)\cos\theta]\; ,
\end{equation}
where $x = 2E_{\nu}/E_{\mu}$ measures the neutrino energy and 
$\hat{s}_{\mu}$ specifies the muon's spin direction.

At the energies best suited for the study of neutrino 
oscillations---tens of GeV, by our current estimates---the muon 
storage ring is compact.  We could build it at one laboratory, pitched 
at a deep angle, to illuminate a laboratory on the other side of the 
globe with a neutrino beam whose properties we can control with great 
precision.  By choosing the right combination of energy and 
destination, we can tune future neutrino-oscillation experiments to 
the physics questions we will need to answer, by specifying the ratio 
of path length to neutrino energy and determining the amount of matter 
the neutrinos traverse.  Although we can use each muon decay only 
once, and we will not be able to select many destinations, we may be 
able to illuminate two or three well-chosen sites from a 
muon-storage-ring neutrino source.  That possibility---added to the 
ability to vary the muon charge, polarization, and energy---may give 
us just the degree of experimental control it will take to resolve the 
outstanding questions about neutrino oscillations.

\textit{The Detector Challenge.} To distinguish oscillations among 
$\nu_{e}$, $\nu_{\mu}$, $\nu_{\tau}$, and a possible fourth, 
``sterile,'' neutrino $\nu_{s}$ that does not experience weak 
interactions, we require a target \textit{cum} detector of several 
kilotonnes---perhaps several tens of kilotonnes---that can
distinguish electrons, muons, and taus, and measure their charges.  
This is a straightforward requirement for the muons, but the 
short-lived ($0.3\hbox{ picosecond}$) $\tau$ and the eager-to-shower 
electron are more difficult to deal with.

\section*{What Can We Learn from the Top Quark?}
Top is a most remarkable particle, even for a quark.\footnote{See
Ref.\ {\protect\cite{cqpt}} for a general introduction.} A single top
quark weighs $175\gevcc$, about as much as an atom of gold.  But
unlike the gold atom, which can be disassembled into 79 protons, 79
electrons, and 118 neutrons, top seems indivisible, for we discern no
structure at a resolution approaching $10^{-18}\hbox{ m}$.  Top's
expected lifetime of about 0.4 yoctosecond ($0.4 \times 10^{-24}\s$)
makes it by far the most ephemeral of the quarks.  The compensation
for this exceedingly brief life is a measure of freedom: top decays
before it experiences the confining influence of the strong
interaction.  In spite of its fleeting existence, the top quark helps
shape the character of the everyday world.

\begin{figure}[tb] 
\centerline{\BoxedEPSF{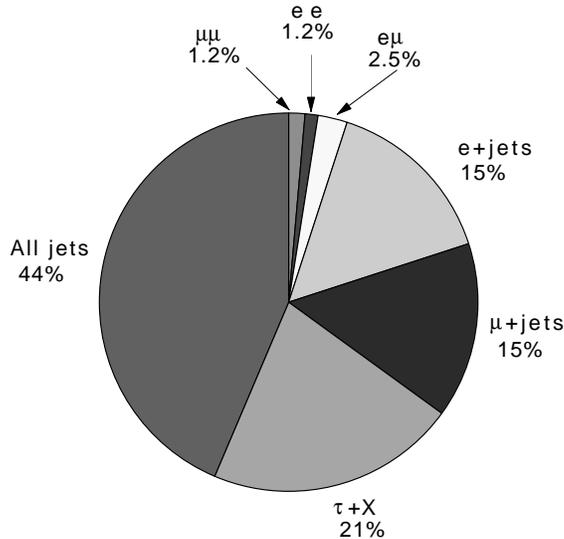  scaled 450}}
\vspace{10pt}
\caption{Branching fractions for prominent decay modes in $t\bar{t}$ 
production.}
\label{fig:tarte}
\end{figure}

The discovery experiments were carried out at Fermilab's Tevatron, in
which a beam of 900-GeV protons collides with a beam of 900-GeV
antiprotons.  Creating top-antitop pairs in sufficient numbers to
claim discovery demanded exceptional performance from the Tevatron,
for only one interaction in ten billion results in a top-antitop pair,
through the reaction
\bothdk{\bar{p}p}{t}{\;\;\bar{t}+\cdots}{W^{-}\bar{b}}{W^{+}b}{ttbar}
Observing traces of the disintegration of top into a $b$-quark and a
$W$-boson, the agent of the weak interaction, required highly capable
detectors and extraordinary attention to experimental detail.  Both
the $b$-quark and the $W$-boson are themselves unstable, with many
multibody decay modes.  The $b$-quark's mean lifetime is about
$1.5\ps$.  It can be identified by a decay vertex displaced by a
fraction of a millimeter from the production point, or by the
low-momentum electron or muon from the semileptonic decays $b
\rightarrow ce\nu$, $b \rightarrow c\mu\nu$, each with branching
fraction about 10\%.  The $W$ boson decays after only $0.3\ys$ on
average into $e\bar{\nu}_{e}$, $\mu\bar{\nu}_{\mu}$,
$\tau\bar{\nu}_{\tau}$, or a quark and antiquark (observed as two jets
of hadrons), with probabilities 1/9, 1/9, 1/9, and 2/3.  The
characteristic modes in which $t\bar{t}$ production can be sought are
shown with their relative weights in Figure \ref{fig:tarte}.
Dilepton events ($e\mu,$ $ee$, and $\mu\mu$) are produced primarily
when both $W$ bosons decay into $e\nu$ or $\mu\nu$.  Events in the
lepton + jets channels $(e,\mu + \hbox{jets})$ occur when one $W$
boson decays into leptons and the other decays through quarks into
hadrons.  Another challenge to experiment is the complexity of events
in high-energy $\bar{p}p$ collisions.  The top and antitop are
typically accompanied by scores of other particles.  The discovery 
experiments scanned $10^{6}$ events per second.

\begin{figure}[tb] 
\centerline{\BoxedEPSF{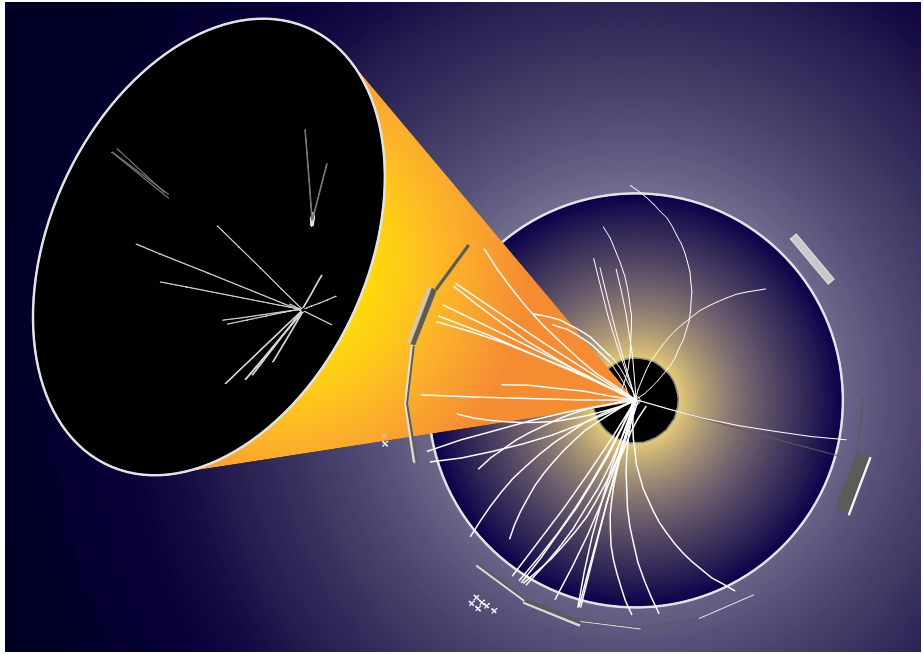  scaled 800}}
\vspace{10pt}
\caption{A pair of top quarks reconstructed in the CDF experiment at 
Fermilab.  Each top decays to a $W$ boson and a $b$ quark.  The tower 
pointing east-southeast in the wide view identifies a positron from $W^{+}$ 
decay; the inset shows displaced decays of two $b$ particles.}
\label{fig:CDFtop}
\end{figure}

\begin{figure}[tb] 
\centerline{\BoxedEPSF{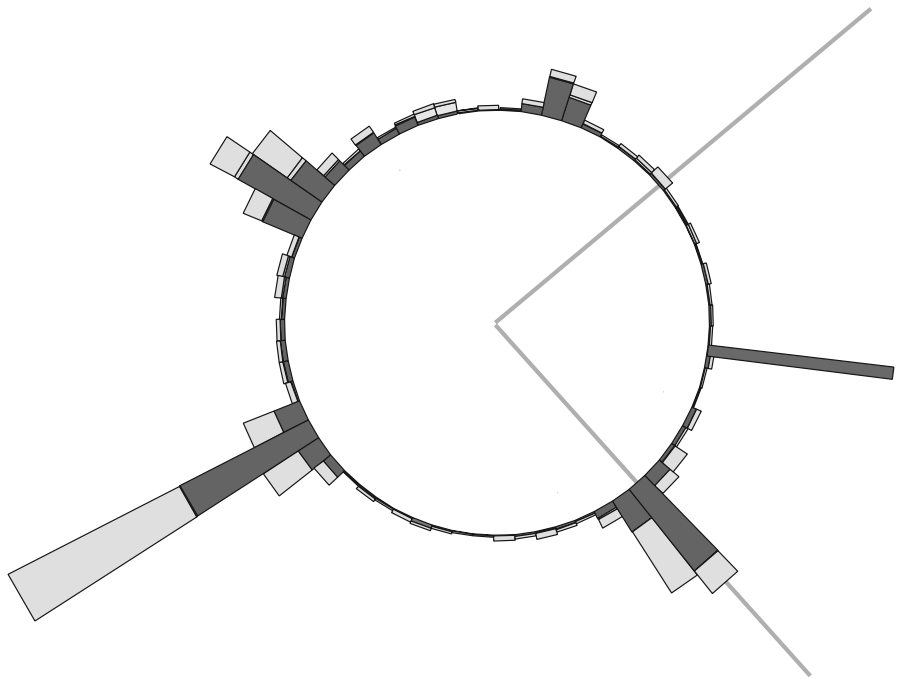  scaled 800}}
\vspace{10pt}
\caption{A pair of top quarks reconstructed in the D\O\ experiment at 
Fermilab.  This end view shows the final decay products: two muons 
(thin lines at northeast and southeast), a neutrino (east), and four 
jets of particles.}
\label{fig:Dzerotop}
\end{figure}

Each detector is an intricate apparatus operated by an 
international collaboration of about 450 physicists.  The 
tracking
devices, calorimeters, and surrounding iron for muon identification 
occupy a volume about three stories high and weigh about 5000 tons.  
The Collider Detector at Fermilab (CDF), a magnetic detector with 
solenoidal geometry, profited from its high-resolution silicon vertex 
detector (SVX) to tag $b$-quarks with good efficiency.  The D\O\ Detector 
(D-Zero) had no central magnetic field, emphasizing instead calorimetric 
measurement of the energies of produced particles.
A top event from the CDF detector, shown in Figure \ref{fig:CDFtop}, 
displays the power of the silicon vertex detector to resolve 
secondary $b$ decays at only a small remove from the production vertex.
The D\O\ event pictured in Figure \ref{fig:Dzerotop} shows one $W$ 
boson reconstructed from a muon plus missing energy, and a second $W$ 
reconstructed from two jets.

\textit{The Detector Challenge.} To select top-quark events with high
efficiency, and to measure them effectively, require the ability to
decide quickly which few of the $10^{7}$ events per second in Tevatron
Run 2 ($10^{8}$ per second at the LHC) are interesting; excellent $b$
tagging, resolving secondary vertices within a few tenths of a
millimeter of the collision point in a hostile high-radiation
environment; efficient lepton identification and measurement of both
electron and muon momenta, even for leptons in jets; calorimetry that
provides good jet-jet invariant mass resolution and a reliable
measurement of missing transverse energy; a hermetic detector.

For the moment, the direct study of the top quark belongs to the
Tevatron.  Early in the next century, samples twenty times greater
than the current samples should be in hand, thanks to the increased
event rate made possible by Fermilab's Main Injector and upgrades to
CDF and D\O. Boosting the Tevatron's energy to $1\tev$ per beam will
increase the top yield by nearly 40\%.  Further enhancements to
Fermilab's accelerator complex are under study.  A decade from now,
the Large Hadron Collider at CERN will produce tops at more than ten
thousand times the rate of the discovery experiments. 
Electron-positron linear colliders or muon colliders may add new
opportunities for the study of top-quark properties and dynamics.  In
the meantime, the network of understanding known as the standard model
of particle physics links the properties of top to many phenomena to
be explored in other experiments.

How can we expect new experiments to extend our knowledge of the top
quark?\footnote{The home page of the Fermilab \textit{thinkshop} on
top-quark physics for Run 2 of the Tevatron, at
\textsf{http://lutece.fnal.gov/thinkshop/}, contains many useful
links.  See also the surveys in Refs.\ \cite{Simmons:1998my} and
\cite{Willenbrock:1998ne}.} Tevatron Run 2, to begin in March 2001, is
to accumulate $2\fb^{-1}$ of integrated luminosity.  We anticipate a
determination of the top mass to $\pm 3\gevcc$ in Run 2, $\pm 1\gevcc$
with $30\fb^{-1}$ in Run 2$^{\mathit{bis}}$ or in LHC experiments.  
Combined with a measurement of the $W$-boson mass to $\pm 40\mevcc$,
this measurement will make it possible to infer the Higgs-boson mass
with increased precision.  It is now possible to begin asking how precisely 
top fits the profile of anticipated properties in its production and 
decay.  It should be possible to determine the $t
\to b W$ coupling in single-top production: $\pm 10\%$ in Run 2, $\pm
5\%$ in Run 2$^{\mathit{bis}}$.  One of the exploratory goals of the
new round of experiments will be to search for $t\bar{t}$ resonances,
rare decays, and deviations from the expected pattern of top decays. 
Finally, it may prove possible to begin to probe the ephemeral
lifetime of top.  On the theoretical front, the large mass of 
top encourages us to think that the two problems of mass may be 
linked at the electroweak scale.

\subsection*{Top Matters!}
It is popular to say that 
top quarks were produced in great numbers in the 
fiery cauldron of the Big Bang some fifteen billion years ago, disintegrated 
in the merest fraction of a 
second, and vanished from the scene until my colleagues learned to create 
them in the Tevatron.  That 
would be reason enough to care about top: to learn how it helped sow the 
seeds for the primordial universe that evolved into our world of diversity 
and change.  But it is not the whole story; it invests the top quark with a 
remoteness that veils its importance for the everyday world.

The real wonder is that here and now, every minute of every day, the top 
quark affects the world around us.  Through the uncertainty principle of 
quantum mechanics, top quarks and antiquarks wink in and out of an 
ephemeral presence in our world.  Though they appear virtually, fleetingly, 
on borrowed time, top quarks have real effects.

Quantum effects make the coupling strengths of the fundamental 
interactions---appropriately normalized analogues of the 
fine-structure constant $\alpha$---vary with the energy scale on 
which the coupling is measured.  The fine-structure constant itself 
has the familiar value $1/137$ in the low-energy (or long-wavelength) 
limit, but grows to about $1/129$ at the mass of the $Z^{0}$ boson, 
about $91\gevcc$.  Vacuum-polarization effects make the effective 
electric charge increase at short distances or high energies.

\begin{figure}[tb]
\centerline{\BoxedEPSF{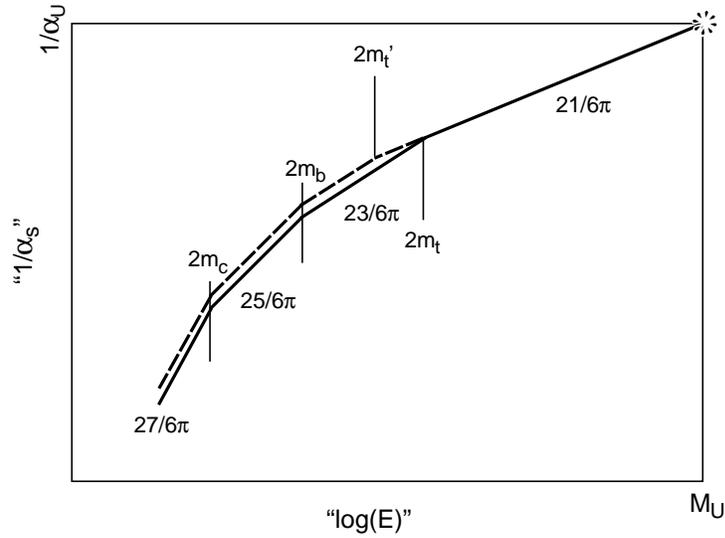  scaled 600}}
\vspace{10pt}
\caption{Two evolutions of the strong coupling constant $\alpha_{s}$. 
A smaller value of the top-quark mass leads to a smaller value of
$\alpha_{s}$.
}
\label{fig:courant}
\end{figure}

In unified theories of the strong, weak, and electromagnetic
in\-ter\-ac\-tions, all the coupling ``constants'' take on a common
value, $\alpha_U$, at some high energy, $M_U$.  If we adopt the point
of view that $\alpha_{U}$ is fixed at the unification scale, then the
mass of the top quark is encoded in the value of the strong coupling
$\alpha_s$ that we experience at low energies.\footnote{All the
important features emerge in an $SU(5)$ unified theory that contains
the standard-model gauge group $SU(3)_{c}\otimes SU(2)_{L} \otimes
U(1)_{Y}$.  The final result is more general.} Assuming three
generations of quarks and leptons, we evolve $\alpha_s$ downwards in
energy from the unification scale.\footnote{The strategy is explained
in Ref.\ {\protect\cite{GQW}}.} The leading-logarithmic behavior is
given by
\begin{equation}
1/\alpha_s(Q) = 1/\alpha_U + \frac{21}{6\pi}\ln(Q/M_U)\;\; ,
\end{equation} for $M_U > Q > 2 m_t$.  The positive coefficient 
$+21/6\pi$ means that the strong coupling constant 
$\alpha_{s}$ is smaller at high energies than at low energies.  This 
behavior---opposite to the familiar behavior of the electric 
charge---is the celebrated property of asymptotic freedom.
In the interval between $2m_t$ and $2m_b$, the 
slope $(33-2n_{\!f})/6\pi$ (where $n_{\!f}$ is the number of active quark 
flavors) steepens to $23/6\pi$, and then increases by 
another $2/6\pi$ at every quark threshold.  At the boundary $Q=Q_n$ 
between effective field theories with $n-1$ and $n$ active flavors, the 
coupling constants $\alpha_s^{(n-1)}(Q_n)$ and $\alpha_s^{(n)}(Q_n)$ must 
match.  This behavior is 
shown by the solid line in Figure \ref{fig:courant}.
The dotted line in Figure \ref{fig:courant} shows how the evolution of
$1/\alpha_s$ changes if the top-quark mass is reduced.  A smaller top
mass means a larger low-energy value of $1/\alpha_s$, so a smaller
value of $\alpha_s$.

To discover the dependence of $\Lambda_{\hbox{{\footnotesize QCD}}}$ upon the top-quark 
mass, we calculate $\alpha_s(2m_t)$ 
evolving up from low energies and down from the unification scale, and match:
\begin{equation}
1/\alpha_U  +  {\displaystyle \frac{21}{6\pi}}\ln(2m_t/M_U) =  
 1/\alpha_s(2m_c) - {\displaystyle \frac{25}{6\pi}}\ln(m_c/m_b) 
 -{\displaystyle \frac{23}{6\pi}}\ln(m_b/m_t)   .
 \end{equation} Identifying
 \begin{equation}
 1/\alpha_s(2m_c) \equiv {\displaystyle\frac{ 
 27}{6\pi}}\ln(2m_c/\Lambda_{\hbox{{\footnotesize QCD}}})\;  ,
\end{equation}  we find that 
\begin{equation}
	\Lambda_{\hbox{{\footnotesize QCD}}}=e^{\displaystyle -6\pi/27\alpha_U} 
	\left(\frac{M_U}{1 \gev}\right)^{\!21/27} 
	\left(\frac{2m_t\cdot 2m_b\cdot 2m_c}{1\gev^{3}}\right)^{\!2/27}\gev \;\; .
	\label{blank}
\end{equation}  

Thanks to QCD, we have learned that the dominant contribution to the 
light-hadron masses is not the masses of the quarks of which they are 
constituted, but the energy stored up in confining the quarks in a 
tiny volume.\footnote{An accessible essay on our understanding of hadron 
mass appears in Ref.\ \cite{fwpt}.} Our most useful tool in the 
strong-coupling regime is lattice QCD.  Calculating the light hadron 
spectrum from first principles has been one of the main objectives of 
the lattice program, and important strides have been made recently.  
In 1994, the GF11 Collaboration \cite{ref:GF11}carried out a quenched 
calculation of the spectrum (no dynamical fermions) that yielded 
masses that agree with experiment within 5--10\%, with good 
understanding of the residual systematic uncertainties.  The CP-PACS 
Collaboration centered in Tsukuba has embarked on an ambitious program 
that will soon lead to a full (unquenched) calculation \cite{burk}.

Neglecting the tiny ``current-quark'' masses of the up and down 
quarks, the scale parameter $\Lambda_{\hbox{\footnotesize QCD}}$ is the 
only mass parameter in QCD.  It determines the scale of the 
confinement energy that is the dominant contribution to the proton mass. 
To a good first approximation, 
\begin{equation}
	M_{\hbox{{\footnotesize proton}}} \approx C \Lambda_{\hbox{{\footnotesize QCD}}},
	\label{lattice}
\end{equation}
where the constant of proportionality $C$ is calculable using 
techniques of lattice field theory.
 
We conclude that, in a simple unified theory,
\begin{equation}
	\frac{M_{\hbox{{\footnotesize proton}}}}{1\gev} \propto 
	\left(\frac{m_t}{1\gev}\right)^{2/27} \;\; .
	\label{amazing}
\end{equation}
This is a wonderful result.  Now, we can't use it to
compute the mass of the top quark, 
because we don't know the values of $M_{U}$ and $\alpha_{U}$, and 
haven't yet calculated precisely the constant of proportionality 
between the proton mass and the QCD scale parameter.  Never mind!  The 
important lesson---no surprise to any twentieth-century physicist---is
that the microworld does determine the behavior 
of the quotidian.  We will fully understand 
the origin of one of the most important parameters in the everyday 
world---the mass of the proton---only by knowing the properties of the 
top quark.\footnote{For a fuller development of the influence of 
standard-model parameters on the everyday world, see Ref.\ 
{\protect\cite{18params}}.}  

\section*{What Is the Dimensionality of Spacetime?}
Ordinary experience tells us that spacetime has $3+1$ dimensions.  
Could our perceptions be limited?  That is the question raised 
allegorically in \textit{Flatland,} a Victorian fable published in 
1880 by a British schoolmaster, Edwin Abbott Abbott (1839 -- 1926), 
and still widely available \cite{flatland}.

Like any question that tests our preconceptions and unspoken 
assumptions, ``What is the dimensionality of spacetime?'' is a 
legitimate scientific question, to which we should return from time to 
time.  It is given immediacy by recent theoretical work.
For its internal consistency, string theory requires an additional six 
or seven space dimensions, beyond the $3+1$ dimensions of everyday 
experience.  Until recently it has been presumed that the extra 
dimensions must be compactified on the Planck scale, with a
compactification radius 
\begin{equation}
R_{\mathrm{unobserved}} \simeq \frac{1}{M_{\mathrm{Planck}}} = 
\frac{1}{1.22 \times 10^{19}\gevcc} = 1.6 \times 10^{-35}\m\; .
\end{equation}
Part of the vision of string theory is that what goes on in the small 
curled-up dimensions does affect the everyday world: excitations of 
the Calabi--Yau manifolds determine the fermion spectrum, for 
example.\footnote{For a gentle introduction to the aspirations of 
string theory, see Ref.\ \cite{beegee}.}

\begin{figure}[tb] 
\centerline{\BoxedEPSF{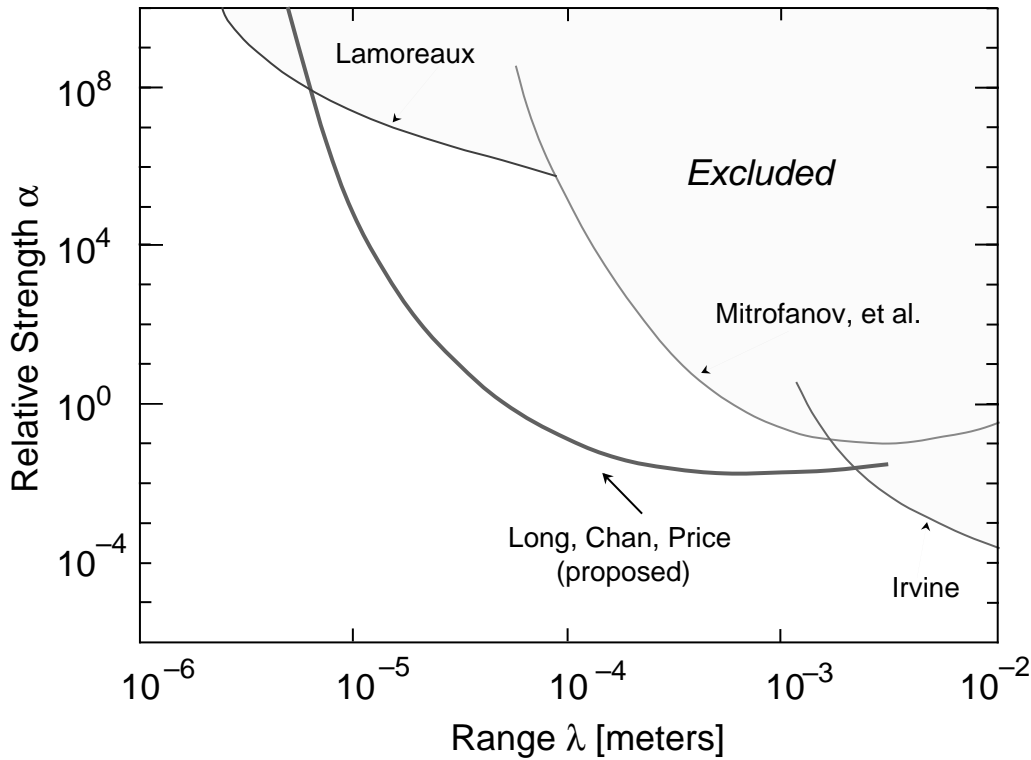 scaled 750}}
\vspace{10pt}
\caption{Experimental limits on the strength $\varepsilon_{\mathrm{G}}$ 
(relative to gravity) versus the range $\lambda_{\mathrm{G}}$ of a 
new long-range force, together with the anticipated sensitivity of a 
new experiment based on small mechanical resonators {\protect\cite{price}}.}
\label{fig:nonNgrav}
\end{figure}

The great gap between the electroweak scale of about $10^{3}\gev$ and the 
Planck scale of about $10^{19}\gev$
gives rise to the hierarchy problem of the  electroweak theory 
\cite{hier}.
The conventional approach to new physics has been to extend the 
standard model to understand why the electroweak scale (and the mass 
of the Higgs boson) is so much smaller than the Planck scale.  A novel 
approach that has evolved over the past two years is instead to 
\textit{change gravity} to understand why the Planck scale is so much 
greater than the electroweak scale \cite{EDbiblio}.  Now, experiment 
tells us that gravitation closely follows the Newtonian force law down 
to distances on the order of $1\mm$.  Let us parameterize deviations 
from a $1/r$ gravitational potential in terms of a relative strength 
$\varepsilon_{\mathrm{G}}$ and a range $\lambda_{\mathrm{G}}$, so that
\begin{equation}
V(r) = - \int dr_{1}\int dr_{2} 
\frac{G_{\mathrm{Newton}}\rho(r_{1})\rho(r_{2})}{r_{12}} \left[ 1+ 
\varepsilon_{\mathrm{G}}\exp(-r_{12}/\lambda_{\mathrm{G}}) \right]\; ,	
\end{equation}
where $\rho(r_{i})$ is the mass density of object $i$ and $r_{12}$ is 
the separation between body 1 and body 2.  Elegant experiments that 
study details of Casimir and Van der Waals forces imply bounds on 
anomalous gravitational interactions, as shown in Figure 
\ref{fig:nonNgrav}.  Below about a millimeter, the constraints on 
deviations from Newton's inverse-square force law deteriorate rapidly, 
so we are free to consider changes to gravity even on a small but 
macroscopic scale.

That is precisely the possibility raised by the interesting new idea 
that some extra dimensions of spacetime might be---relatively 
speaking---large.\footnote{For an enthusiastic overview, see Marcus 
Chown, ``Five and Counting \ldots,'' \textit{New Scientist} 
\textbf{160,} No.  2157 (24 October 1998), 
\textsf{http://www.newscientist.com/ns/981024/fifth.html}.}
The new idea is to consider that the 
$SU(3)_{c}\otimes SU(2)_{L}\otimes U(1)_{Y}$ standard-model gauge 
fields, plus needed extensions, reside on $3+1$-dimensional branes, 
not in the extra dimensions, but that gravity can propagate into the 
extra dimensions. How does this hypothesis change the picture?  The 
dimensional analysis (Gauss's law, if you like) that relates Newton's 
constant to the Planck scale changes.  If gravity propagates not only 
in the $3+1$ dimensions of Minkowski space, but also in $n$ 
extra dimensions with radius $R$, then
\begin{equation}
    G_{\mathrm{Newton}} \sim M_{\mathrm{Planck}}^{-2} \sim 
    M^{\star\,-n-2}R^{-n}\; ,
    \label{eq:gauss}
\end{equation}
where $M^{\star}$ is gravity's true scale.  The correlation between 
$M^{\star}$ and the size $R$ of the $n$ large extra dimensions is 
given in Table \ref{tbl:ed} for some representative cases.
\begin{table}[tb]
    \caption{Radius $R$ of $n$ large extra dimensions for low values 
    of the scale $M^{\star}$ of gravity.}
    \renewcommand{\baselinestretch}{1.3}
    \begin{tabular}{ccccc}
         & $n=1$ & $n=2$ & $n=3$ & $n=6$  \\
        \hline
        $M^{\star}=1\tevcc$ & $10^{13}\m$ & $10^{-3}\m$ & $10^{-8}\m$ & 
        $10^{-14}\m$  \\
        $M^{\star}=10\tevcc$ & $10^{10}\m$ & $10^{-5}\m$ & $10^{-10}\m$ & 
        $10^{-15}\m$  \\
        $M^{\star}=100\tevcc$ & $10^{7}\m$ & $10^{-7}\m$ & $10^{-12}\m$ & 
        $10^{-16}\m$  \\
        \hline
    \end{tabular}
    \renewcommand{\baselinestretch}{1}
    \label{tbl:ed}
\end{table}
Could the extra dimensions be quasimacroscopic?  One large extra 
dimension seems excluded, since gravity within the solar system obeys 
Newton's force law in three (not more) spatial 
dimensions.\footnote{The semimajor axis of Pluto's orbit is about $6 
\times 10^{12}\m$.} Notice that if we boldly take $M^{\star}$ to be as 
small as $1\tevcc$, then the scaling law \eqn{eq:gauss} requires the 
radius of the extra dimensions to be smaller than about $1\mm$, for $n 
\ge 2$.\footnote{Other observational constraints 
{\protect\cite{shrock}} suggest that it is more prudent to choose 
$M^{\star} \approx 10\hbox{ - }100\tevcc$, for which at least three 
large extra dimensions are required to alter gravity on the millimeter 
scale.} Deviations on the millimeter scale are allowed by current 
knowledge of gravity on short distances, but will be challenged soon.  

\begin{figure}[tb] 
\centerline{\BoxedEPSF{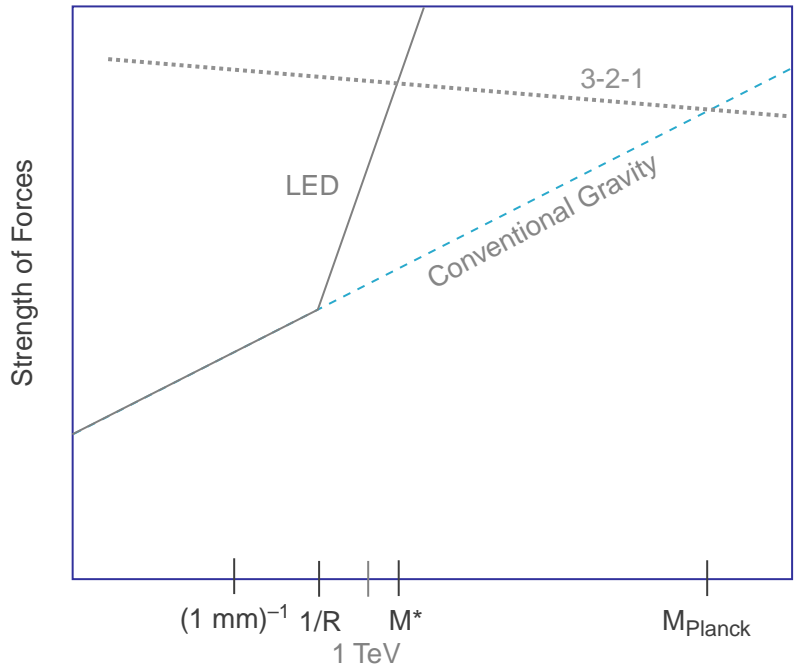 scaled 750}}
\vspace{10pt}
\caption{One of these extrapolations (at least!) is false.}
\label{fig:false}
\end{figure}

If we use the four-dimensional force law to 
extrapolate the strength of gravity from low energies to high, we find 
that gravity becomes as strong as the other forces on the Planck 
scale, as shown by the dashed line in Figure \ref{fig:false}.  If the 
force law changes at an energy $1/R$, as the large-extra-dimensions 
scenario suggests, then the forces are unified at an energy 
$M^{\star}$, as shown by the solid line in Figure \ref{fig:false}.  
What we know as the Planck scale is then a mirage that results from a 
false extrapolation: treating gravity as four-dimensional down to 
arbitrarily small distances, when in fact---or at least in this 
particular fiction---gravity propagates in $3+n$ spatial dimensions.  
The Planck mass is an artifact, given by $M_{\mathrm{Planck}} = 
M^{\star}(M^{\star}R)^{n/2}$.

Extra dimensions seen by standard-model particles cannot be larger 
than about $1\tev^{-1} \approx 10^{-19}\m$, or we would already have 
probed them at LEP2, HERA, and the Tevatron.  We will see in a moment 
that future collider experiments can examine two or more large extra 
dimensions with a low gravitational (string) scale $M^{\star}$.

Although the idea that extra dimensions are just around the 
corner---either on the submillimeter scale or on the TeV scale---is 
preposterous, it is not ruled out by observations.  For that reason 
alone, we should entertain ourselves by entertaining the consequences.  
Any particle can radiate a graviton into extra dimensions.  An 
extradimensional graviton is only gravitationally coupled, and so will 
not interact in the detector: the gravitons go off into extra 
dimensions and are lost.  Their signature is missing energy, 
$\slashii{E}_{T}$.  These processes, individually tiny, may be 
observable because the number of excitable modes is very large.  Many 
authors \cite{shrock} have considered the gravitational excitation of a tower of 
Ka\l uza--Klein modes in the extra dimensions, which would give rise 
to a missing (transverse) energy signature in collider experiments 
\cite{smaria}.  We call these excitations \textit{provatons,} after 
the Greek word $\pi\rho\acute{o}\beta\alpha\tau o$ for a sheep in a 
flock.  For proton-antiproton collisions, the experimental signatures 
include $\bar{p}p \rightarrow {\mathrm{jet}} + 
\slashii{E}_{T}\;\;\hbox{(parton + graviton)}$, and $\bar{p}p 
\rightarrow \ell^{+}\ell^{-} + \slashii{E}_{T}\;\;(\ell^{+}\ell^{-} + 
\hbox{graviton)}$.
	
\textit{The Detector Challenge.} To establish a ${\mathrm{jet}} + 
\slashii{E}_{T}$ signature at a hadron collider, we require a hermetic 
detector with well-controlled $\slashii{E}_{T}$ tails to establish and 
quantify the missing-energy signature; highly efficient rejection of 
cosmic-ray and accidental triggers; the ability to reject triggers 
originating from jet mismeasurements; and control over physics 
backgrounds, including $Z^{0} + \mathrm{jets}$ and $W^{\pm} + 
\mathrm{jets}$. The experimental sensitivity to provatons 
depends on collider energy, luminosity, and species, and the 
number of large extra dimensions.

A representative analysis of the constraints that may be inferred from 
anomalous single photon production at $e^{+}e^{-}$ colliders and from 
monojet production at hadron colliders if no signal is seen is shown in Table 
\ref{tbl:mpp}.  
\begin{table}[tb]
    \caption{Sensitivities to large extra dimensions, expressed as 95\% 
    CL upper limits on the radius $R$ of $n$ extra dimensions (from 
    Ref.\ {\protect\cite{mpp}}).}
    \renewcommand{\baselinestretch}{1.3}
    \begin{tabular}{ccccc}
		 & $n=2$ & $n=4$ & $n=6$  \\
		\hline
		LEP2 & $4.7 \times 10^{-4}\m$ & $1.9 \times 10^{-11}\m$ & $6.9 
		\times 10^{-14}\m$  \\
		Tevatron Run 1 & $1.1 \times 10^{-3}\m$ & $2.4 \times 10^{-11}\m$ & 
		$5.8 \times 10^{-14}\m$  \\
		\hline
		Tevatron Run 2 & $3.9 \times 10^{-4}\m$ & $1.4 \times 10^{-11}\m$ & 
		$4.0 \times 10^{-14}\m$  \\
		Large Hadron Collider & $3.4 \times 10^{-5}\m$ & $1.9 \times 
		10^{-12}\m$ & $6.1 \times 10^{-15}\m$  \\
		1-TeV polarized linear collider & $1.2 \times 10^{-5}\m$ & $1.2 \times 
		10^{-12}\m$ & $6.5 \times 10^{-15}\m$  \\
		\hline
    \end{tabular}
    \renewcommand{\baselinestretch}{1}
    \label{tbl:mpp}
\end{table}
If a missing-energy signal is found, it will be a challenge to 
distinguish an extradimensional signal from the classic 
$R$-parity--conserving signature for supersymmetry.  
We should be so lucky!

\subsection*{The Outlook for Extra Dimensions}
We are only beginning to explore the possible implications of 
almost-accessible extra dimensions.  Among the fascinating new worlds
to explore are the Randall--Sundrum mechanism for localizing gravity
in an extra dimension (in the vicinity of a brane or domain wall) \cite{lisa}, 
and the speculation of Arkani-Hamed and Schmaltz that the 
fermion mass hierarchy reflects fermion wave packets separated in an 
extra dimension \cite{martin}.  The characteristic missing-energy 
signature of a Ka\l uza--Klein excitation will be hard to distinguish 
from other new physics.  We need to think in more detail about the 
backgrounds and optimal search techniques.  The search for large extra 
dimensions reinforces the informative metaphor of a collider and its 
detectors as an ultramicroscope.  Are extra dimensions large enough to 
see?  It is also interesting to consider what might we observe above 
the threshold for exciting extra dimensions.  Although the basic 
concepts that underlie these speculations have a sound basis in string 
theory, most of the scenarios put forward so far have more to do with 
storytelling than with theoretical rigor.  However, it is plain that 
our inability to disprove at once the outlandish idea of large extra 
dimensions is a measure of the potential for experimental surprises, 
and an indication of Nature's capacity to amaze us!
\section*{Final Thoughts}
To a great degree, the progress of particle physics has followed from 
progress in accelerator science and instrumentation.  There is no 
substitute for experiment, and experiment requires inventions in both 
hardware and software and continuous innovation in analysis.  The 
slogan, ``Yesterday's sensation is today's calibration and tomorrow's 
background,''\footnote{I believe that this formulation is due to V. 
L. Telegdi.} embodies both the challenge and the opportunity of 
advances in experimental technique.

In the middle of the revolution we are experiencing---indeed,
making---in our conception of Nature, when we deal with fundamental
questions about our world, including
\begin{verse}
    \textit{What are the symmetries of Nature, and how are they hidden 
        from us?}
    
        \textit{Are the quarks and leptons composite?}
    
        \textit{Are there new forms of matter, like the superpartners 
                suggested by supersymmetry?}
    
        \textit{Are there more fundamental forces?}
        
        \textit{What makes an electron an electron, a neutrino a neutrino, and a top
                quark a top quark?}
    
        \textit{What is the dimensionality of spacetime?}
\end{verse}    
we cannot advance without new instruments that extend our senses and
allow us to create---and understand---new experience far beyond the
realm of everyday human experience.
	
I wish you a productive two weeks in Istanbul, and hope that what you 
learn about instrumentation, the spirit of experimentation, and the 
habits of mind necessary to make detectors act as reliable extensions 
of our senses will open new horizons for you and for science.

\section*{Acknowledgments}
It is a pleasure to thank our Istanbul hosts for their 
energetic and delightful hospitality.  I commend the Panel on 
Instrumentation Innovation and Development of the International 
Committee for Future Accelerators for their sponsorship of this series 
of instrumentation schools, and salute the leaders of the 
instrumentation courses for their infectious enthusiasm.  On behalf of 
the students, I thank the great laboratories of particle physics, 
particularly CERN and Fermilab, for providing much of the apparatus 
that makes possible the hands-on experience that gives this school its 
special character.  Finally, I would like to express my thanks to the 
students who came to Istanbul from many interesting parts of the globe 
and whose lively curiosity made my visit memorable.  I hope I may have 
the opportunity to welcome many of you to Fermilab in the future.

\end{document}